# Manipulating fractional Shapiro steps in twisted cuprate Josephson junctions

Yuying Zhu[1,4#*], Heng Wang[2,3#], Ding Zhang[1,3,4*], and Qi-Kun Xue[1,2,3*]

[1]Beijing Academy of Quantum Information Sciences, Beijing 100193, China

[2]Southern University of Science and Technology, Shenzhen 518055, China

[3]State Key Laboratory of Low Dimensional Quantum Physics and Department of Physics, Tsinghua University, Beijing 100084, China

[4]Hefei National Laboratory, Hefei 230088, China

\# These authors contributed equally.

*Email: zhuyy@baqis.ac.cn

dingzhang@mail.tsinghua.edu.cn

qkxue@mail.tsinghua.edu.cn

# Abstract

High-quality Josephson junctions made of twisted cuprate superconductors offer unprecedented opportunities in addressing fundamental problems and realizing next-generation superconducting devices at relatively high temperatures. Whether or not the twisted cuprates possess high-temperature topological superconductivity remains an outstanding issue. Here, we tackle this problem via an in-depth study of the key predicted feature—half-integer Shapiro steps. We show that half-integer Shapiro steps do occur in samples at a twist angle of $45°$ but are unstable with thermal cycling. Interestingly, fractional steps can be introduced by training the sample with a small magnetic field or annealing with a large electrical current, attesting to a tunable current-phase relation (CPR) in twisted cuprates. We extend the current annealing to realize fractional steps with odd denominators too. Furthermore, half-integer steps can be induced in the regime that is well beyond the expectation of topological superconductivity, favoring an alternative mechanism involving trapped vortices. Our results not only caution the direct association of half-integer Shapiro steps with the exotic mechanism but also open a distinct pathway toward a Josephson junction with electrically tunable CPR at high temperatures.

**Introduction**

The Josephson junction under a constant bias $V$ carries a rapidly oscillating current at a frequency of $2eV/h$. This ac Josephson effect usually manifests itself as current steps at finite voltage values—Shapiro steps—in the $I$-$V$ characteristic of the junction under microwave irradiations [1]. The standard Shapiro steps situate at: $V_n = \frac{hf}{2e}n$, where $n$ is an integer and $f$ is the microwave frequency. The quantized values of $V_n$ make Josephson junctions the voltage standard in metrology [2, 3]. Apart from the conventional sequence of integer steps, the ac Josephson effect may exhibit fractional steps at $V_{n,m} = \frac{hf}{2e}\frac{n}{m}$, where $n$ and $m$ are both integers and $n \neq m$. Such an unusual sequence was observed earlier by Clarke in a superconductor-normal metal-superconductor (SNS) junction [4]. Similar subharmonics appeared in other SNS junctions [5-8], including grain boundary junctions made of high-temperature superconductors [9, 10]. They can be attributed to the synchronization between certain multiples of the Josephson frequency ($2meV/h$) and higher harmonics of the microwave [11]—an epitome of nonlinear dynamical systems hosting the so-called devil's staircases [12, 13]. The microscopic mechanism, depending on the specific situations, may involve the presence of multiple Andreev reflections [5, 6], micro-shorts [14], or synchronized motion of vortices in a wide junction [10].

Recently, non-trivial topological physics has been proposed as a distinct source for the fractional ac Josephson effect [15, 16]. The system of interest consists of two *d*-wave superconductors stacked vertically (along the *c*-axis) with a twist angle of 45 degrees [17] (For simplicity, disorder effects and competing charge ordering are neglected). This twisted system is theoretically predicted to spontaneously break time-reversal symmetry and host high-temperature topological superconductivity. One predicted feature is the emergence of Shapiro steps at half-integers (1/2, 3/2, etc.) due to the co-tunneling of Cooper pairs. Experimentally, twisted cuprates of high interface quality have been realized for testing this proposal but the results remain contradictory. The twisted bicrystal of Bi$_2$Sr$_{2-x}$La$_x$CuO$_{6+y}$ (Bi-2201) showed no fractions in its Fiske

steps—resonances between the ac Josephson effect and cavity modes of the junction [18]. Yet another experiment with $Bi_2Sr_2CaCu_2O_{8+x}$ (Bi-2212) observed directly the indications of half-integer Shapiro steps [19] between the integral steps at a twist angle of 44.6°. There, the mechanism was claimed to be different from vortex induced effect [10], because no external magnetic field was applied and the standard Fraunhofer pattern observed in other fabricated samples (at 44.9° and 46.3°) suggested the absence of parallel junctions. Still, half-integer Shapiro steps were observed so far only in a single sample of twisted cuprates and at a relatively high temperature of 70 K. The contradicting results between different groups urgently call for clarifications because further proposals utilizing the topological superconductivity have been made. These proposals include high-temperature Majorana physics [16,20-22], flowermon quantum computations [23,24], as well as charge-4$e$ superconductivity [25].

Here, we investigate the ac Josephson effect in several twisted Bi-2212 junctions with different twist angles. We obtain Shapiro steps in all samples under microwave irradiations at temperatures across a wide range (10-70 K). Half-integer steps indeed appear in 45°-twisted junctions but they are sensitive to the thermal cycling process. Interestingly, they can reappear by either training the sample with a small magnetic field or applying a large current to the sample during the cool-down. We show that the current annealing can even introduce features at fractions of 1/3 and 2/3 with the absence of steps at 1/2 and 3/2, clearly beyond the scope of present theory on topological superconductivity. Moreover, we observe prominent half-integer Shapiro steps in twisted junctions with twist angles of 30° and 40° at relatively high temperatures after an electrical current training process. Our results indicate that extrinsic effects can be responsible for the half-integer Shapiro steps in the twisted cuprates. The developed method can effectively and reversibly alter the current-phase relation (CPR), demonstrating the advantage of twisting in realizing novel superconducting devices.

## Results

### Half-integer Shapiro steps and its removal by thermal cycling

We start by showing sharply contrasting behaviors of ac Josephson effects in the same sample (device A). This sample, with the optical image shown in Fig. 1a, possesses a twist angle $\theta$ of 45°. The data in Fig. 1b-f and Fig. 1h-l are obtained at zero magnetic field (see Methods). Figure 1b shows fully overlapping superconducting transition behaviors of the junction resistance over multiple thermal cycles. The superconducting transition temperature $T_c$ is determined to be 88 K (pointed out by an arrow in Fig. 1b), demonstrating high-quality of the junction without oxygen loss throughout the fabrication [26]. Each curve is taken from a temperature above the transition temperature $T_c$ (110-140 K) to 1.6 K. It indicates that the transport quality of this sample stays constant throughout the measurement. Figure 1 c and d compare the tunneling current-voltage ($I$-$V$) characteristics at 5 K of device A in two cool-downs. The Josephson critical current $I_c$ is obviously larger for the data in Fig. 1d.

The difference in $I_c$ is correlated with the different responses in the ac Josephson effect (Fig. 1e, f, h-l). Here we plot the derivative $dI/dV$—tunnel conductance—or $dV/dI$—tunnel resistance—as a function of the junction voltage ($V$)/current ($I$) and the microwave power ($P$). In Fig. 1e, h-k, the Shapiro steps manifest themselves in the color-coded plot of $dI/dV(V,P)$ as vertical stripes with high intensities (orange color). In the corresponding plot of $dV/dI(I,P)$, they emerge as dark-colored stripes (Fig. 1f, l). In the first cool-down, apart from the main sequence at integer steps, we indeed observe additional high intensity regions at half integers (indicated by red triangles) in Fig. 1e. They manifest as additional splitting of the resistive region in the $dV/dI(I,P)$ plot of Fig. 1f. These features suggest the presence of half-integer Shapiro steps [19]. In Fig. 1g, we theoretically simulate the occurrence of fractional

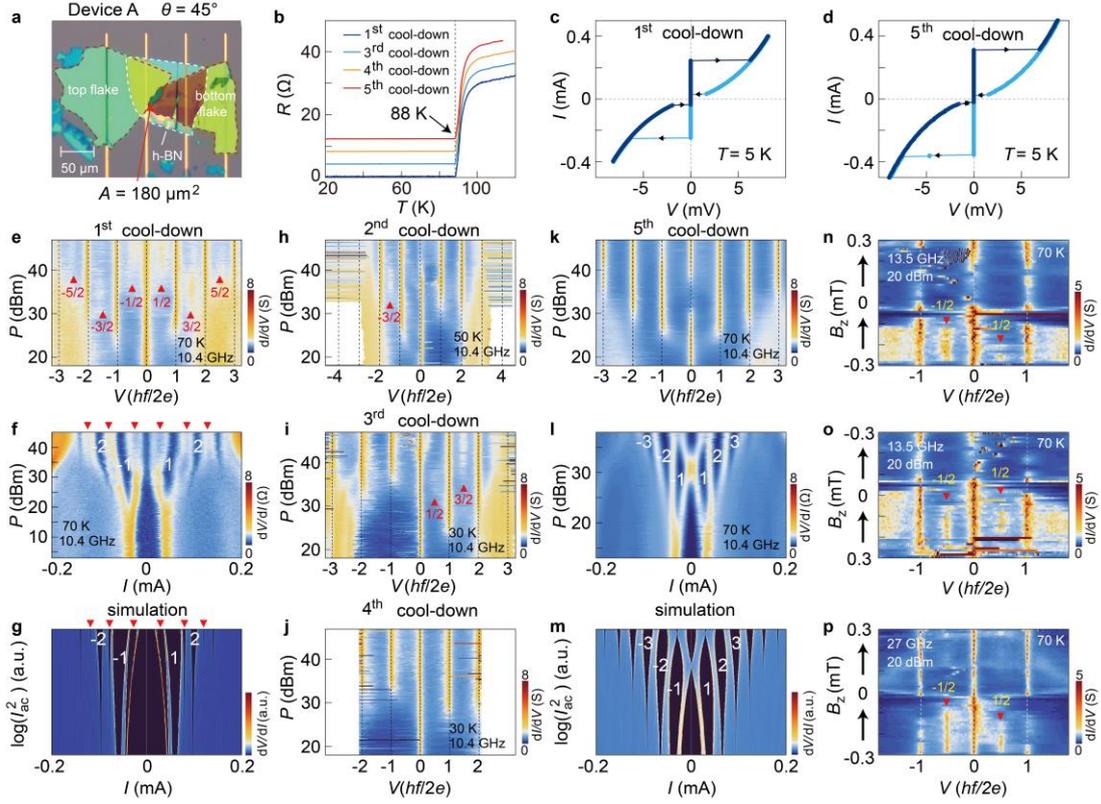

*Figure 1: Half-integer Shapiro steps in a 45°-twisted cuprate junction.* All experimental data are from device A. **a,** Optical image of device A. **b,** Temperature dependent junction resistance measured in different cool-downs. Curves are vertically offset for clarity. **c, d,** I-V characteristics in two separate sample cool-downs. Arrows indicate the sweeping directions. **e, h, i, j, k,** Color-coded plot of tunnel conductance $dI/dV$ as a function of bias voltage $V$ (in unit of $hf/2e$, where $f$ is the microwave frequency) and microwave power $P$ in different cool-downs. **f, l,** Color-coded plot of tunnel resistance $dV/dI$ as a function of $I$ and $P$ in the first and fifth cool-downs, respectively. **g, m,** theoretically simulated evolution of the Shapiro steps by either including (**g**) or excluding (**m**) the second harmonic term in the CPR of a resistively shunted junction. Red triangles in **e-i** and **n-p** mark the Shapiro steps at half integers. **n, o, p,** $dI/dV$ as a function of $V$ and the out-of-plane magnetic field $B_z$ under microwave irradiations with two frequencies (13.5 GHz and 27 GHz). The microwave power is 20 dBm. Black arrows on the left side indicate the sweeping directions of the magnetic field.

Shapiro steps [27] by considering a CPR that contains both first and second harmonic terms ($J_{c1} \sin \phi$ and $J_{c2} \sin 2\phi$). To mimic the situation of Fig. 1f, we select a ratio of $J_{c2}/J_{c1}$ to be about 1. The coexistence of both the first and second harmonics is different from the theoretical expectation in which only the second harmonic term remains at the twist angle of 45°.

The presence of the second harmonic term seems to validate the prediction of topological superconductivity in twisted cuprates. However, instead of a single observation, we carry out repeated measurements while thermal cycling the sample to a temperature above $T_c$ and back. We observe that the occurrence of fractional steps depends sensitively on the thermal cycling. Figure 1h-k presents data obtained in the subsequent cool-downs. Only one fractional step at -3/2 persists after the first thermal cycling (Fig. 1h). In the third cool-down, fractional steps occur at 1/2 and 3/2 in the positive side at 30 K (Fig. 1i). In the fourth and fifth cool-downs, however, fractional steps fully vanish (Fig. 1j, k, l). Figure 1m presents the corresponding theoretical modelling [27] of the situation in Fig. 1l. Here, the CPR contains only the first harmonic term. Structural instability and oxygen loss are less likely to account for this sensitivity because we only raise the temperature to about 140 K, still in the cryogenic regime. We surmise that unintentionally trapped vortices may cause the half-integer steps. This scenario was employed to explain the fractional steps observed in grain boundary junctions [28, 29]. Notably, ref. [28] reported the similar metastable behavior of the half-integer steps with thermal cycling. The trapped vortices could be generated by the remanent magnetic field, which is about 0.2 mT in our case (Methods).

**Half-integer Shapiro steps induced by magnetic field training**

In Fig. 1n-p, we show that half-integer Shapiro steps reappear in device A at 70 K under a rather small magnetic field. We plot the tunneling conductance as a function of bias voltage and magnetic field, while fixing $f$ and $P$. Only integer steps at $\pm 1$ can be seen in the section when sweeping the magnetic field from 0 to 0.3 mT (Fig 1n) or from 0 to -0.3 mT (Fig. 1o). By contrast, both integer steps at $\pm 1$ and half-integer steps at $\pm 1/2$ appear when sweeping the magnetic field from a finite value back to 0. In Fig. 1p, we confirm this hysteretic behavior at another microwave frequency. Such a field-sweeping dependence is reminiscent to the hysteretic field response of the magnetization of a superconductor with flux pinning centers [30], again pointing to the link between the emergence of half-integer Shapiro steps and trapped vortices. In general, both the perpendicular and in-plane magnetic fields can induce half-integer Shapiro steps. Our results suggest that the trapped flux in the sample connects the Josephson vortex in the plane and the Abrikosov vortex that points out-of-plane, similar to the configuration proposed in ref. [31].

**Fractional Shapiro steps of a metastable state**

Figure 2 collects data from device B ($\theta = 45°$, $T_c = 88$ K). Here we observe unusual behaviors of the ac Josephson effect at an intermediate temperature (~ $0.5T_c$). Figure 2 a and b show basic tunneling characterizations of device B without microwave irradiations. We cool down the sample from 150 K to 1.6 K ($B = 0$ T) and then measure $I$-$V$ characteristics at each temperature point in the warm-up sequence (from 5 to 90 K). The temperature dependence of the critical current $I_c$ is consistent with our previous observation [26], which is distinctly different from that expected for a co-tunneling process. Figure 2c-e present the two-dimensional (2D) scan of $dI/dV$ as a function of $V$ and $f$ at three consecutively reached temperatures in a single cool-down. High-intensity stripes fall on straight lines (dashed) starting from (0,0). They show slopes $dV/df$ that are consistent with $nh/2e$, where $n = \pm 1, \pm 2, \pm 3 \ldots$, indicating the existence of integer Shapiro steps. The clear separation between the high intensity stripes at 70 K and 30 K suggests the absence of fractional Shapiro steps. By contrast, the data at 50 K (Fig. 2d) shows stronger fluctuations, with high intensity in $dI/dV$ occasionally situating between the integer Shapiro steps.

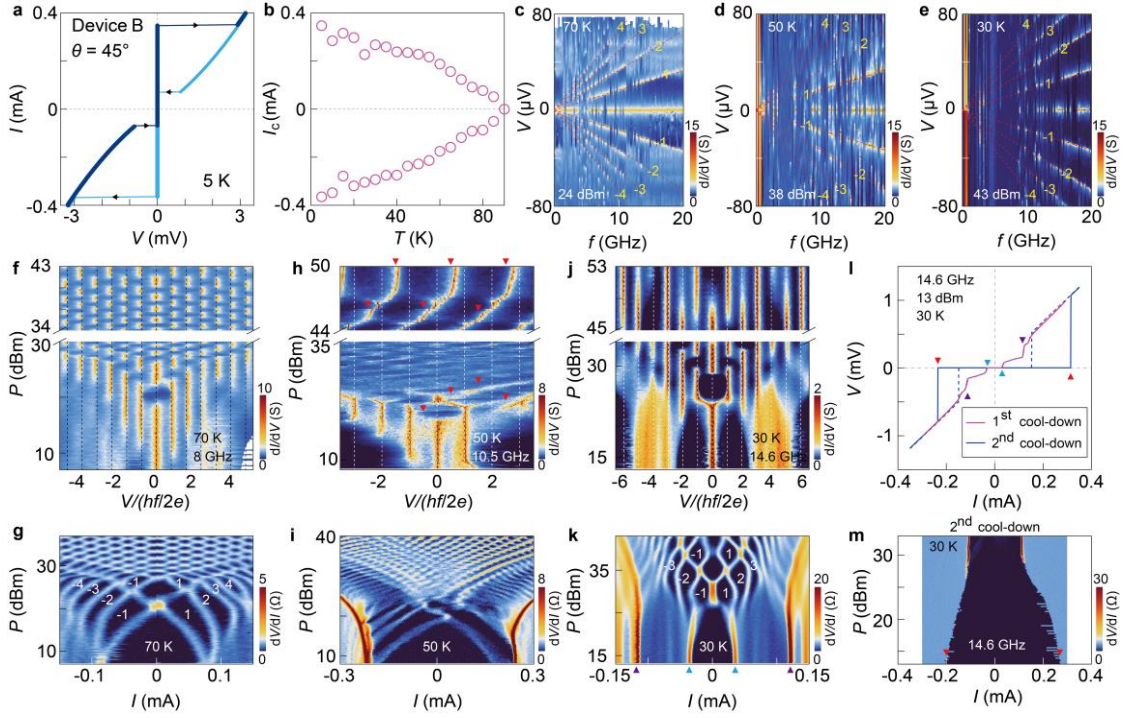

*Figure 2: Fractional Shapiro steps in a second 45°-twisted cuprate junction.* All data are from device B. **a,** $I$-$V$ characteristics at 5 K (microwave off). **b,** Josephson critical current as a function of temperature (microwave off). **c, d, e,** $dI/dV$ as a function of $V$ and $f$ at three consecutively reached temperature points (in the same cool-down). Dashed lines show the calculated positions for the integer Shapiro steps with increasing $f$ (numbers mark the sequence). **f, g, h, i, j, k,** $dI/dV$ as a function of $V$ and $P$ (**f, h, j**) or $dV/dI$ as a function of $I$ and $P$ (**g, i, k**) at three consecutive temperature points 70 K, 50 K, and 30 K (in the same cool-down). Dashed lines in **f, h,** and **j** indicate the positions of integer Shapiro steps. Numbers in **g** and **k** indicate the sequence of Shapiro steps. Red triangles in **h** mark the seemingly fractional Shapiro steps. **l,** Comparison between $I$-$V$ characteristics obtained in two cool-downs. Triangles indicate the jumps in the current sweeps. **m,** $dV/dI$ as a function of $P$ and $I$ at 30 K in the second cool-down.

To elucidate the origin of the instability, we analyze $dI/dV(V,P)$ or $dV/dI(I,P)$ obtained at 70 K, 50 K, and 30 K in a single cool-down. Figure 2f and g demonstrate that there exist only integer Shapiro steps at 70 K. The rich evolution of the stripes as a function of $P$ agrees with the standard theory of ac Josephson effect [32]. Interestingly, the subsequent microwave measurement at 50 K (Fig. 2h, i) yields anomalous evolution. In particular, the high intensity region shows continuous evolution between the integer steps, giving rise to sloped stripes instead of vertically aligned stripes in the $dI/dV(V,P)$ plot (Fig. 2h). This behavior is prominent at high $P$, where the local maxima of $dI/dV$ seem to locate at half-integers. If these seemingly fractional steps are related to topological superconductivity, one would expect that they become more prominent at lower temperatures. However, further cooling down device B to 30 K and irradiating the microwave give rise to only well aligned integer Shapiro steps, as shown in Fig. 2j, k. It suggests that the seemingly fractional steps at 50 K may have a different origin than the intrinsic mechanism, which predicts a larger topological gap thus more prominent half-integer steps at lower temperature [16].

A closer look of the data at 30 K (Fig. 2k) reveals that there exist multiple jumps toward the normal state (indicated by blue and purple triangles). In Fig. 2l, we compare two $I$-$V$ characteristics, both obtained from device B at 30 K (microwave on). We refer to the behavior with more than two jumps in either the positive or the negative sweep as state-1. The data showing the standard two switching events in each sweeping direction corresponds to state-2. Figure 2m shows that microwave irradiation at state-2 results in only gradual suppression of the critical current, without causing any Shapiro steps. The absence of Shapiro steps for state-2 reflects the fact that the critical current density is large at this temperature such that the Josephson plasma frequency greatly exceeds the frequency of the applied microwave [33]. In general, state-1 and state-2 represent dramatically different responses to microwave. We point out that state-1 is reached after sweeping current at high temperatures (70 K, 50 K, etc.), whereas state-2 is approached by cooling directly from 150 K ($>T_c$) to 30 K without applying any large current to the sample at intermediate temperatures. It is therefore likely that the large current applied at 70 K or 50 K before reaching 30 K introduces trapped vortices to the sample. The additional jumps in the $I$-$V$ characteristics can be attributed to the depinning of vortices at the corresponding current. The vortices may undergo thermal activation at 50 K, generating the unusual behavior in the ac Josephson effect.

**Fractional Shapiro steps induced by current annealing**

In Fig. 3, we present the ac Josephson effect of device B measured after taking a current annealing process, corroborating the link between a large electrical current and the emergence of fractional Shapiro steps. Figure 3a illustrates the annealing protocol. We cool down the sample from a temperature well above $T_c$ (150 K, for instance) to 10 K while passing a dc current $I_b$ —in the range of $3I_c(10\text{ K})$ to $6I_c(10\text{ K})$—through the junction. After reaching 10 K, we withdraw the large current and measure the ac Josephson effect at consecutively increased temperatures. In Fig. 3b to 3d, we show the 2D scan of $dI/dV(V,P)$ after current annealing with $I_b = 1.0$ mA. At 30 K, prominent fractional Shapiro steps emerge at $\pm 2/3$ (Fig. 3b). They become less discernible when raising the temperature to 50 K (Fig. 3c), presumably due to the thermal activation of vortices. At 70 K, the sample recovers the state with only integer Shapiro steps. Similar results can be obtained after taking the current annealing process with increasing $I_b$ (Fig. 3e-h). We also identify signatures of fractional Shapiro steps at $\pm 1/3$.

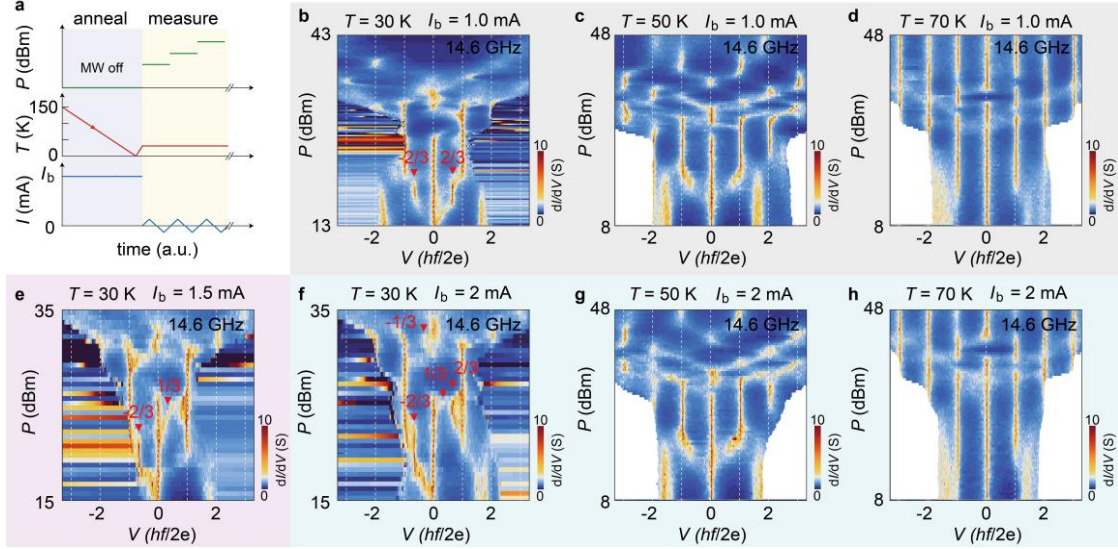

***Figure 3: Fractional Shapiro steps induced by current annealing. a,*** *current annealing protocol. A large bias current $I_b$ is passed through the sample during the cool-down without irradiating microwave (MW off).* ***b, c, d,*** *Color-coded plot of $dI/dV$ as a function of $P$ and $V$ (in unit of $hf/2e$) for device B after current annealing with $I_b$ = 1 mA at three consecutively reached temperature points: 30 K, 50 K, 70 K (in the warm-up direction). Dashed lines indicate the integer Shapiro steps.* ***e,*** *Color-coded plot of $dI/dV$ as a function of $P$ and bias voltage $V$ (in unit of $hf/2e$) for device B after current annealing with $I_b$ = 1.5 mA at 30 K.* ***f, g, h,*** *Color-coded plot of $dI/dV$ as a function of $P$ and $V$ (in unit of $hf/2e$) for device B after current annealing with $I_b$ = 2 mA at three consecutively reached temperature points: 30 K, 50 K, 70 K (in the warm-up direction). Red triangles in **b**, **e**, and **f** mark the observed fractional Shapiro steps.*

Passing a large dc current through the sample can induce a self-field exceeding the lower critical field at temperatures close to $T_c$. This self-field may exceed the lower critical field of the junction. Consequently, the self-field generates vortices that get trapped in the twisted junction during the cool-down. We note that our sample with reduced thickness makes it more susceptible to magnetic fluxes, due to the much-increased London penetration depth. The trapped vortices may play a similar role as that of a small magnetic field in a wide junction [10], giving rise to higher harmonics in the current phase relation of the Josephson junction. Such a scenario has been explicitly considered before in high-temperature grain boundary junctions [28]. The modeling of trapped Josephson vortices there yielded nice agreement with the observed fractional Shapiro steps. We comment that the current annealing may result in a different arrangement of the trapped vortices, such that the obtained $dI/dV(V,P)$ plots at 30 K in Fig. 3 are not exactly the same as Fig. 2j. Still, current annealing turns out to be effective in artificially manipulating the CPR of the twisted junction.

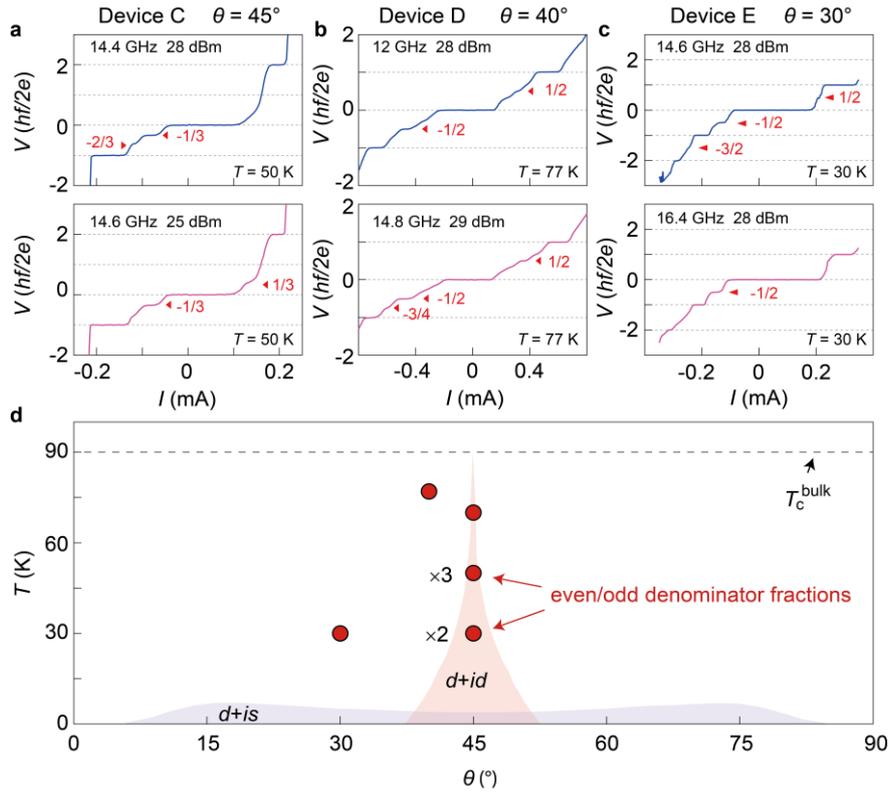

**Figure 4: Fractional Shapiro steps in multiple devices. a, b, c,** $I$-$V$ characteristics for three devices (C, D, E) at the specified microwave frequency and power after applying the current pulses. Red triangles indicate the fractional Shapiro steps. **d,** Summary of the twist angles and temperature points at which fractional Shapiro steps are observed in this study. Shaded regions indicate the theoretically predicted regimes for the d+id-wave and d+is-wave pairing.

In Fig. 4a-c, we collect supporting results from another three samples (device C, $\theta = 45°$; device D, $\theta = 40°$; device E, $\theta = 30°$). Instead of current annealing, we send current pulses directly at a fixed temperature below $T_c$. Immediately after taking this procedure, $I_c$ gets suppressed without applying the microwave and Shapiro steps occur under microwave irradiation. Accompanied by the suppression, fractional

Shapiro steps with either odd or even denominators appear in different samples. Interestingly, we observe fractional Shapiro steps even at 3/4, attesting to the appearance of the devil's staircases [11]. We further comment that the original state of the samples, in terms of a large $I_c$, can be recovered after thermal cycling.

## Discussion

Our results caution the immediate attribution of fractional ac Josephson effects in twisted cuprates to the exotic mechanism such as topological superconductivity. We not only reproduce the previously reported half-integer steps at $\theta = 45°$ at 70 K but also demonstrate their disappearance in the subsequent measurements upon thermal cycling. This metastable property seems to run against either *d+id* or *d+is*-wave pairing being the uniform and stable ground state. One may argue that there exist multiple domains of *d±id* (*d±is*) order parameters, similar to the $p_x \pm ip_y$ domains considered before [34-36]. And half-integer Shapiro steps disappear because domains with opposite chirality cancel out. In this scenario, the current pulse or current training can help form one dominating order parameter from the competing ones. However, in the multi-domain scenario, the critical Josephson current in the absence of microwave would be small if domains host opposite chirality. This is because the currents from different domains tend to cancel each other. The critical Josephson current would reach the maximum value when the chirality becomes the same in different domains. When the chirality becomes the same, one would expect to see prominent half-integer Shapiro steps. In short, half-integer Shapiro steps should appear when irradiating the microwave on the sample in a state with a large Josephson current. This is clearly in contradiction with our experimental observation. For example, we observe half-integer Shapiro steps in device A (Fig. 1e) when the critical current without microwave is small (Fig. 1c) and observe purely integer Shapiro steps in the same device (Fig. 1k) when the critical current is large (Fig. 1d). Also, we realize half-integer Shapiro steps at (30°, 30 K) and (40°, 77 K) of the phase diagram by using the current pulse technique. These data points are far outside the predicted regime (shaded regions in Fig. 4d) [16,

37, 38] for *d*+i*d* or *d*+i*s*-wave pairing. A more stringent theoretical consideration [38] showed that the time-reversal symmetry breaking phase would occupy a narrower angular range than that shown in Fig. 4d. Therefore, it is unlikely that *d*+i*d* or *d*+i*s*-wave pairing lurks at (30°, 30 K) and (40°, 77 K), as the competing ground state, for the current pulse to settle the system into.

We note that only half-integer Shapiro steps were predicted for *d*+i*d* or *d*+i*s*-wave pairing, because they stem from the fact that the tunneling is carried by a net charge of $4e$. In principle, higher harmonics may occur but they should be accompanied by the more dominant second harmonic term [37]. By contrast, we observe odd-denominator fractional steps at $\pm 1/3$ and $\pm 2/3$ but no half-integer steps in the 45°-twisted junction (device B, device C) after current annealing/pulsing (Figs. 3 and 4a). In fact, Shapiro steps at fractions with only odd denominators were reported in a wide Josephson junction under a certain magnetic field [10]. Even in zero nominal magnetic field, half-integer Shapiro steps were observed in high-temperature grain boundary junctions and attributed to trapped vortices [28]. There too, a small critical Josephson current correlated with the occurrence of half-integer Shapiro steps, echoing with our observations. The scenario of trapped vortices is applicable to a broad scope of materials, including the recent example of MgB$_2$ junction [29], in which only *s*-wave pairing exists. This effect is valid if the actual device hosts parallel junctions. We argue that this existence of multiple junctions seems intrinsically related to the small Josephson penetration depth of Bi-2212 (<1 $\mu m$) [18], which is smaller than the physical size of the overlapped area between the two twisted cuprate flakes.

Finally, we discuss the possible implication of this study. We repeatedly observe prominent Josephson tunneling in the optimally doped sample at the twist angle of 45 degrees [18,26]. The ac Josephson effect showing purely integer Shapiro steps in these samples suggests that higher order harmonics (including the second harmonic) play a negligible role. It seriously challenges the attribution of the prominent Josephson tunneling at 45 degrees to the emergence of a second harmonic term. Instead, the results seem to reaffirm our previous conclusion that there exists a simple *s*-wave component in the twisted cuprates. Future experiments may direct toward elucidating the role of super-modulation, charge ordering, disorder and emergent ordering in inducing such a simple *s*-wave component.

In summary, we provide a comprehensive study of the ac Josephson effect in twisted cuprates at different twist angles ($45°$, $40°$, and $30°$) and across a wide temperature range (10 to 70 K). We show that half-integer Shapiro steps can occasionally occur without applying any magnetic field in $45°$-twisted junctions. We further study this unusual phenomenon by employing thermal cycling, magnetic field training and current annealing. The sensitive response of the half-integer Shapiro steps to the above-mentioned variations suggest that they may be a result of trapped vortices. We further demonstrate that current sweeping at temperatures below $T_c$ can induce fractional Shapiro steps, at half-integers and at $\pm 1/3$ and $\pm 2/3$, in the twisted junctions. Such an electrical tuning method works for junctions at $30°$ and $40°$ too. This study exposes the rich evolution of the ac Josephson effect in twisted cuprates, offering a platform for quantitatively analyzing the higher harmonics of the CPR. Our work also outlines a measurement protocol (zero-current cool) for extracting intrinsic properties of sensitive Josephson junctions. The inversion symmetry breaking in the twisted cuprates, together with the induced time-reversal symmetry breaking, can help generate interesting effects such as the superconducting diode effect [39].

**Methods**

High-quality and optimally doped Bi-2212 single crystals were employed in our experiments. We fabricated twisted Josephson junctions by using the on-site cold stacking method described previously [26]. The entire process was completed in a glovebox with argon atmosphere ($H_2O$ < 0.01 ppm, $O_2$ < 0.01 ppm). All junctions are protected by capping a h-BN flake such that the devices can be taken out of the storing glovebox for measurements multiple times within a time span of a year.

Transport measurements were performed in two different closed-cycle Helium-4 systems. One of them was equipped with a superconducting magnet and can be cooled to a base temperature of 1.6 K. We intentionally warmed up the solenoid to a temperature well above the transition temperature. We then cooled down the system and made sure that no current was sourced to the solenoid magnet throughout the measurements. The other cryogenic system essentially consisted of a cold finger directly on the second cooling stage of the Gifford-McMahon (GM) cryocooler. It offered a base temperature of 3.5 K and was equipped with a detachable solenoid wound by copper wires for applying a small magnetic field in a precisely controlled fashion (via a current source meter).

Two microwave sources were employed: one with a maximum frequency of 20 GHz and the other one up to 40 GHz. For the cryogenic system with the superconducting magnet, we sent the microwave through a coaxial cable (diameter: 0.86 mm) with Beryllium copper as both the inner and outer conductors. The specified attenuation is 6.58 dB/m at 10 GHz, 9.44 dB/m at 20 GHz, 14.5 dB/m at 40 GHz. The coaxial cable has a total length of about 0.8 m. The low temperature side of the coaxial cable was cut open with the inner conductor protruded, serving as the antenna.

We employed the fluxgate magnetometer with a resolution of 0.1 nT to measure the remanent magnetic field in our cryogenic systems. We determined the remanent field in the center of the superconducting magnet to be 0.2 mT in the vertical direction

(along the rotational axis of the solenoid) and 0.02 mT in the horizontal direction. The stray field at the same horizontal level but outside the cryogenic system was about 0.09 mT. For the one directly on the GM cryocooler, the remanent field at the sample stage was measured to be 0.04 mT in the vertical direction and 0.02 mT in the horizontal direction. This remanent field arises from the earth's magnetic field. We in total measured the ac Josephson effect in five samples with the twist angle of 45 degrees in the system with the superconducting magnet. Only one device (device A) showed half-integer Shapiro steps in the cool-down without current annealing or current pulsing. Device B, which had a larger junction area, showed no half-integer Shapiro steps at 70 K and 30 K. The rare occurrence of half-integer steps suggests that the remanent field only occasionally induces trapped vortices into the device. In the other cryogenic system, in which the remanent field was determined to be about 0.04 mT, we did not see half-integer Shapiro steps in samples that were cooled down without current annealing. Therefore, the earth's magnetic field seems to play a negligible role.

The potential misalignment of the magnetic field for the measurement in Fig. 1n, o, p may stem from the limited precision of the machined parts and the slight tilting of the sample relative to the sample stage. We estimate the former misalignment to be about 0.5 degree if assuming the machining inaccuracy to be 1 mm (one order of magnitude higher than the typical value claimed by the manufacturer) and taking the fact that the diameter of the solenoid is 125 mm. The slight tilting of the sample is estimated to be 1.9 degree, assuming a drastic height difference of the glue used for fixing the sample to be 0.1 mm (the thickness of the glue) and taking the fact that the width of the substrate is 3 mm. With the applied perpendicular magnetic field of 0.3 mT (Fig. 1n, o, p), the total misalignment of 2.5 degrees would give rise to an in-plane component of 0.013 mT. This tiny field is buried in the earth's magnetic field that we measure at the sample stage.

**Data availability**

The data for plotting the figures in the main text have been deposited in: https://doi.org/10.57760/sciencedb.20150.

**Acknowledgements**

We thank Haoyu Wang, Steffen Wahl, Jurgen H. Smet for the valuable input in the initial stage of the project. We thank Genda Gu for providing us with the Bi-2212 crystals. This work is financially supported by the Ministry of Science and Technology of China [Grant No. 2022YFA1403100 (DZ, QKX)]; National Natural Science Foundation of China [Grants No. 52388201 (QKX), T2425009 (DZ), 12141402 (YZ), 12274249 (DZ), 92477101 (YZ)]; Innovation Program for Quantum Science and Technology [Grants No. 2021ZD0302600 (YZ), 2021ZD0302400 (DZ)]; the China Postdoctoral Science Foundation [Grant No. GZB20240294 (HW), 2024M751287 (HW)].

**Author contributions**

Y. Z. and H. W. fabricated the devices and carried out the transport measurements. All authors analyzed the data, discussed the results and wrote the paper.

**References:**

[1] S. Shapiro. Josephson Currents in Superconducting Tunneling: The Effect of Microwaves and Other Observations. *Phys. Rev. Lett.* **11**, 80 (1963).

[2] B. F. Field, T. F. Finnegan, and J. Toots. Volt Maintenance at NBS via 2e/h: A New Definition of the NBS Volt. *Metrologia* **9**, 155-166 (1973).

[3] K. v. Klitzing. Essay: Quantum Hall Effect and the New International System of Units. *Phys. Rev. Lett.* **122**, 200001 (2019).

[4] J. Clarke. Finite-Voltage Behavior of Lead-Copper-Lead Junctions. *Phys. Rev. B* **4**, 2963 (1971).

[5] P. Dubos, H. Courtois, O. Buisson, and B. Pannetier. Coherent Low-Energy Charge Transport in a Diffusive S-N-S Junction. *Phys. Rev. Lett.* **87**, 206801 (2001).

[6] J. C. Cuevas, J. Heurich, A. Martín-Rodero, A. Levy Yeyati, and G. Schön. Subharmonic Shapiro Steps and Assisted Tunneling in Superconducting Point Contacts. *Phys. Rev. Lett.* **88**, 157001 (2002).

[7] J. Basset, M. Kuzmanović, P. Virtanen, T. T. Heikkilä, J. Estève, J. Gabelli, C. Strunk, and M. Aprili. Nonadiabatic dynamics in strongly driven diffusive Josephson junctions. *Phys. Rev. Research* **1**, 032009(R) (2019).

[8] A. Iorio, A. Crippa, B. Turini, S. Salimian, M. Carrega, L. Chirolli, V. Zannier, L. Sorba, E. Strambini, F. Giazotto, and S. Heun. Half-integer Shapiro steps in highly transmissive InSb nanoflag Josephson junctions. *Phys. Rev. Research* **5**, 033015 (2023).

[9] E. A. Early, A. F. Clark and K. Char. Half-integral constant voltage steps in high-$T_c$ grain boundary junctions. *Appl. Phys. Lett.* **62**, 3357–3359 (1993).

[10] D. Terpstra, R. P. J. IJsselsteijn and H. Rogalla. Subharmonic Shapiro steps in high-$T_c$ Josephson junctions. *Appl. Phys. Lett.* **66**, 2286–2288 (1995).

[11] Yu. M. Shukrinov, S. Yu. Medvedeva, A. E. Botha, M. R. Kolahchi and A. Irie. Devil's staircases and continued fractions in Josephson junctions. *Phys. Rev. B* **88**, 214515 (2013).

[12] E. Ben-Jacob, Y. Braiman, R. Shainsky and Y. Imry. Microwave-induced "Devil's Staircase" structure and "chaotic" behavior in current-fed Josephson junctions. *Appl. Phys. Lett.* **38**, 822–824 (1981).

[13] I. Sokolović, P. Mali, J. Odavić, S. Radošević, S. Yu. Medvedeva, A. E. Botha, Yu. M. Shukrinov and J. Tekić. Devil's staircase and the absence of chaos in the dc- and ac-driven overdamped Frenkel-Kontorova model. *Phys. Rev. E* **96**, 022210 (2017).

[14] L.C. Ku, H. M. Cho, S. W. Wang. The origin of the half-integral constant voltage steps in high-$T_c$ grain-boundary junction, *Physica C: Superconductivity* **243**, 187 (1995).

[15] Z. S. Yang, S. S. Qin, Q. Zhang, C. Fang, J. P. Hu. π/2-Josephson junction as a topological superconductor. *Phys. Rev. B* **98**, 104515 (2018).

[16] O. Can, T. Tummuru, R. P. Day, I. Elfimov, A. Damascelli, M. Franz. High-temperature topological superconductivity in twisted double-layer copper oxides. *Nat. Phys.* **17**, 519 (2021).

[17] Y. Zhu, M. Liao, Q. Zhang, H.-Y. Xie, F. Meng, Y. Liu, Z. Bai, S. Ji, J. Zhang, K. Jiang, R. Zhong, J. Schneeloch, G. Gu, L. Gu, X. Ma, D. Zhang, and Q.-K. Xue, Presence of s-wave pairing in Josephson junctions made of twisted ultrathin $Bi_2Sr_2CaCu_2O_{8+\delta}$ flakes. *Phys. Rev. X* **11**, 031011 (2021).

[18] H. Wang, et al. Prominent Josephson tunneling between twisted single copper oxide planes of $Bi_2Sr_{2-x}La_xCuO_{6+y}$. *Nat. Commun.* **14**, 5201 (2023).

[19] S. Y. F. Zhao, et al. Time-reversal symmetry breaking superconductivity between twisted cuprate superconductors. *Science* **382**, 1422–1427 (2023).

[20] G. Margalit, B. Yan, M. Franz and Y. Oreg. Chiral Majorana modes via proximity to a twisted cuprate bilayer. *Phys. Rev. B* **106**, 205424 (2022).

[21] A. Mercado, S. Sahoo and M. Franz. High-Temperature Majorana Zero Modes. *Phys. Rev. Lett.* **128**, 137002 (2022).

[22] Y.-X. Li and C.-C. Liu. High-temperature Majorana corner modes in a d + id superconductor heterostructure: Application to twisted bilayer cuprate superconductors. *Phys. Rev. B* **107**, 235125 (2023).

[23] V. Brosco, G. Serpico, V. Vinokur, N. Poccia and U. Vool. Superconducting Qubit Based on Twisted Cuprate Van der Waals Heterostructures. *Phys. Rev. Lett.* **132**, 017003 (2024).

[24] H. Patel, V. Pathak, O. Can, A. C. Potter, and M. Franz. d-mon: A transmon with strong anharmonicity based on planar c-axis tunneling junction between d-wave and

s-wave superconductors. *Phys. Rev. Lett.* **132**, 017002 (2024).

[25] Y.-B. Liu, J. Zhou, C. Wu and F. Yang. Charge-4e superconductivity and chiral metal in 45°-twisted bilayer cuprates and related bilayers. *Nat. Commun.* **14**, 7926 (2023).

[26] Y. Zhu, H. Wang, Z. Wang, S. Hu, G. Gu, J. Zhu, D. Zhang and Q.-K. Xue. Persistent Josephson tunneling between $Bi_2Sr_2CaCu_2O_{8+x}$ flakes twisted by 45° across the superconducting dome. *Phys. Rev. B* **108**, 174508 (2023).

[27] Seoane Souto, R., Leijnse, M., Schrade, C., Valentini, M., Katsaros, G. and Danon, J. Tuning the Josephson diode response with an ac current. *Phys. Rev. Res.* **6**, L022002 (2024).

[28] Early, E. A., Steiner, R. L., Clark, A. F. and Char, K. Evidence for parallel junctions within high-Tc grain-boundary junctions. *Phys. Rev. B* **50**, 9409 (1994).

[29] Yin, D., Cai, X., et al. Half-integer Shapiro steps in MgB2 focused He ion beam Josephson junctions. *Chin. Phys. B* **33**, 087404 (2024).

[30] Poole, C. P., Farach, H. A., Creswick, R. J. & Prozorov, R. 5 - Magnetic Properties. in Superconductivity (Second Edition) (eds. Poole, C. P., Farach, H. A., Creswick, R. J. & Prozorov, R.) 113 – 142 (Academic Press, Amsterdam, 2007). doi: https://doi.org/10.1016/B978-012088761-3/50027-2.

[31] Ghosh, S., Patil, V., et al. High-temperature Josephson diode. *Nat. Mater.* **23**, 612 (2024).

[32] Russer, P. Influence of Microwave Radiation on Current-Voltage Characteristic of Superconducting Weak Links. *Journal of Applied Physics* **43**, 2008–2010 (1972).

[33] Oya, G., Terada, A., Takahashi, N., Irie, A. & Hashimoto, T. Shapiro Step Response of the Surface Intrinsic Josephson Junctions in $Bi_2Sr_2CaCu_2O_y$ Single Crystals at Elevated Temperatures. *Japanese Journal of Applied Physics* **44**, L491 (2005).

[34] Kidwingira, F., Strand, J. D., Van Harlingen, D. J. and Maeno, Y. Dynamical Superconducting Order Parameter Domains in $Sr_2RuO_4$. *Science* **314**, 1267 (2006).

[35] Fernández Becerra, V. and Milošević, M. V. Multichiral ground states in mesoscopic p-wave superconductors. *Phys. Rev. B* **94**, 184517 (2016).

[36] Yasui, Y., Lahabi, K., Becerra, V. F. et al. Spontaneous emergence of Josephson

junctions in homogeneous rings of single-crystal Sr$_2$RuO$_4$. *npj Quantum Mater.* **5**, 21 (2020).

[37] Tummuru, T., Plugge, S. and Franz, M. Josephson effects in twisted cuprate bilayers. *Phys. Rev. B* **105**, 064501 (2022).

[38] Song, X.-Y., Zhang, Y.-H. and Vishwanath, A. Doping a moiré Mott insulator: A t-J model study of twisted cuprates. *Phys. Rev. B* **105**, L201102 (2022).

[39] Wang, H., Zhu, Y., et al. Quantum superconducting diode effect with perfect efficiency above liquid-nitrogen temperature. *arXiv:2509.24764*.